\documentclass[twocolumn,prl,aps,showpacs,superscriptaddress,floatfix]{revtex4-1}
\usepackage[latin1]{inputenc}
\usepackage{bm}
\usepackage{multirow}
\usepackage{amssymb}
\usepackage{amsbsy}
\usepackage{amsmath}
\usepackage{stmaryrd}
\usepackage{graphicx}
\usepackage{placeins}
\usepackage{ulem}
\usepackage[pdftex]{color}

\newcommand{\Le}{\left}
\newcommand{\Ri}{\right}
\newcommand{\nn}{\nonumber}
\newcommand{\f}{\frac}

\newcommand{\ra}{\rangle}
\newcommand{\la}{\langle}

\newcommand{\eq}[1]{\begin{align}#1\end{align}}

\newcommand{\na}{n_\alpha}
\newcommand{\avgna}{\langle n_\alpha \rangle}

\begin{document}

\title{Many-body localization characterized from a one-particle perspective}

\author{Soumya Bera}
\affiliation{Max-Planck-Institut f\"ur Physik komplexer Systeme, 01187 Dresden, Germany}
\author{Henning Schomerus}
\affiliation{Department of Physics, Lancaster University, LA1 4YB Lancaster, United Kingdom}
\affiliation{Max-Planck-Institut f\"ur Physik komplexer Systeme, 01187 Dresden, Germany}
\author{Fabian Heidrich-Meisner}
\affiliation{Department of Physics and Arnold Sommerfeld Center for Theoretical Physics,
Ludwig-Maximilians-Universit\"at M\"unchen, 80333 M\"unchen, Germany}
\author{Jens H.\ Bardarson}
\affiliation{Max-Planck-Institut f\"ur Physik komplexer Systeme, 01187 Dresden, Germany}
\date{\today}
\begin{abstract}
We show that the one-particle density matrix $\rho$ can be used to characterize the interaction-driven many-body localization transition in closed fermionic systems.
The natural orbitals (the eigenstates of $\rho$) are localized in the many-body localized phase and spread out when one enters the delocalized phase,
while the occupation spectrum (the set of eigenvalues of $\rho$) reveals the distinctive Fock-space structure of the  many-body eigenstates, exhibiting a step-like discontinuity in the localized phase.
The associated one-particle occupation entropy is small in the localized phase and large in the delocalized phase, with diverging fluctuations at the transition.
We analyze the inverse participation ratio of the natural orbitals and find that it is independent of system size in the localized phase.
\end{abstract}

\maketitle
{\it Introduction.}
While the theory of noninteracting disordered systems is well developed~\cite{kramer93,evers08}, the possibility of a localization transition in closed interacting systems has only recently been firmly established~\cite{BAA,Mirlin,oganesyan07,Znidaric:2008cr,pal10,Berkelbach:2010ib,Monthus:2010gd,bardarson12,Huse:2013bw, Serbyn:2013he, Imbrie:2014vo, Vosk:2014jl,Pekker:2014bj,kjaell14, Laumann:2014ju,Andraschko:2014bw,chandran14,DeLuca:2014ch,Lazarides2014,Ponte2015,Agarwal:2015cu}.
This many-body localization (MBL) transition occurs at finite energy densities and is not a conventional thermodynamic transition~\cite{altman14,nandkishore15}.
Instead, it can be understood as a dynamical phase transition, associated with the emergence of a complete set of local conserved quantities in the localized phase, which thus behaves as an integrable system~\cite{serbyn13,vosk13,huse14,Ros:2015ib,Chandran:2015cw}.
This restricts the entanglement entropy of the eigenstates to an area law~\cite{Bauer:2013jw}, in contrast to the volume law predicted by the eigenstate thermalization hypothesis for the ergodic delocalized phase~\cite{Deutsch:1991ju,Srednicki:1994dl,Rigol:2008bf}.
At the localization transition, the fluctuations of the entanglement entropy diverge~\cite{kjaell14,Grover:2014wm}.
The effects of MBL are also observed in the dynamics following, for example,  a global quench from a product state, wherein dephasing between the effective degrees of freedom leads to a characteristic logarithmic growth of the entanglement entropy~\cite{bardarson12,Znidaric:2008cr,Serbyn:2013he}.
These features comprise a much richer set of signatures than in the context of noninteracting systems, for which, in the spirit of one-parameter scaling, the notion of a localization length based on single-particle wave functions generally suffices \cite{kramer93,evers08}.

In view of the rich phenomenology of many-body localization it is natural to ask, both from a fundamental point of view as well as for the interpretation of experimental data~\cite{demarco,schreiber15}, to which extent (if at all) the MBL transition can be detected and characterized from a single-particle perspective.
Here we show that a rather complete characterization of many-body localization is indeed possible based on the eigenvalues (occupations) and eigenstates (natural orbitals) of the one-particle density matrix.
The one-particle density matrix was originally introduced by Onsager and Penrose to extend the notion of a Bose-Einstein condensate to interacting systems~\cite{onsager}.
Importantly, the natural orbitals take a Bloch form in translationally invariant systems, providing a true many-body generalization of the Bloch theorem~\cite{koch}.
This naturally suggests studying the effect of disorder, as in recent studies of localization of thermalized hard-core bosons in a quasi-periodic potential~\cite{nessi11,Gramsch:2012bl}.
However, so far no connection to many-body localization has been made.

We are further motivated to consider the one-particle density matrix because it naturally focusses on the dynamics of one particle in the presence of all the others, without the need to resort to a mean-field theory or to sacrifice particle indistinguishability.
As we will see, this perspective retains sufficient information to capture the genuine many-body aspects that set many-body localization apart from Anderson localization transitions in noninteracting systems.
In particular, the occupations sharply reorganize themselves from being close to either zero or one in the localized phase to being in between these extremal values in the delocalized phase, thus reflecting a delocalization transition in Fock space that corresponds to a mixing of product states.
It follows that in the localized phase the occupation spectrum develops a step-like discontinuity, similar to a Fermi-liquid.
The associated one-particle occupation entropy is large and proportional to the system size in the delocalized phase, corresponding to the volume law of thermal states, while in the localized phase it is small.
The dynamics of one particle in the effective bath of the others thus provides complementary information to the dynamics of a spatially confined region in the effective bath of its surrounding.
In addition, we show that the transition leaves direct signatures in the natural orbitals, which are localized in the many-body localized phase and spread out over the system when one enters the delocalized phase.
We show that the inverse participation ratio (IPR) of the natural orbitals depends on the system size in the delocalized phase, while it is independent of system size in the localized phase.

{\it Model and method.} We consider spinless fermions in one dimension with a nearest-neighbor repulsion and diagonal disorder, described by the Hamiltonian
\eq{
H = t\sum_{i=1}^L &\Le\lbrack -\frac{1}{2} (c^{\dagger}_i c_{i+1} + \mathrm{h.c.}) +  \epsilon_i \Le(n_i-\f{1}{2}\Ri) \Ri.  \nn \\
 & \Le. + V \Le(n_i-\f{1}{2}\Ri) \Le(n_{i+1}-\f{1}{2}\Ri) \Ri\rbrack
.
\label{eq:ham}
}
Here $c_i^{\dagger}$ creates a fermion on site $i=1,2,\ldots,L$ and $n_i = c_i^{\dagger}c_i $ is the associated number operator.
Energies are expressed in units of the nearest-neighbor hopping constant $t$, so that $V$ is a dimensionless measure of the strength of the nearest-neighbor repulsive interactions.
The diagonal disorder is introduced via a box distribution of the onsite potentials $\epsilon_i \in \lbrack -W, W \rbrack$.
We study this system using exact diagonalization at finite sizes $L=10,12,14$ ($10^5$ disorder realizations), $L=16$ ($10^3$ realizations) and $L=18$ ($500$ realizations), imposing periodic boundary conditions and fixing the overall occupation at half filling (number of particles $N=L/2$).
We mainly focus on the energy region around the band center, $\varepsilon = 1$ where $\varepsilon = 2(E - E_\text{min})/(E_\text{max}-E_\text{min})$ with $E_\text{max}$ and $E_\text{min}$ the maximum and minimum energy for each disorder realization, and take the $6$ eigenstates closest to this energy. This energy corresponds to infinite temperature in the thermodynamic limit.
%
At the fixed interaction strength $V=1$, the critical disorder strength $W_c$ is found to take values in the range between three and four~\cite{pal10,Berkelbach:2010ib,DeLuca:2013ba,BarLev:2015co,luitz15}.

Given a many-body eigenstate  $| \psi_n \rangle$ of the Hamiltonian~\eqref{eq:ham}, the one-particle density matrix is  defined as
\begin{equation}
\rho_{ij} =\langle \psi_n | c^{\dagger}_i c_j| \psi_n \rangle.
\label{eq:rho}
\end{equation}
The natural orbitals $| \phi_{\alpha} \rangle$ with $\alpha = 1,2,\ldots, L$, are obtained by diagonalizing $\rho$,
\begin{equation}
\rho | \phi_{\alpha} \rangle = n_{\alpha} | \phi_{\alpha} \rangle,
\label{eq:rhoEV}
\end{equation}
which delivers a basis of single-particle states.
The eigenvalues $n_{\alpha}$ are interpreted as occupations, with $\sum_{\alpha=1}^L n_{\alpha}={\rm tr}\,\rho = N$ equal to the total number of particles in the system.
We order the natural orbitals by descending occupation, $n_1\geq n_2\geq\ldots \geq n_L$.

%
\begin{figure}[tb]
\includegraphics[width=0.95\columnwidth]{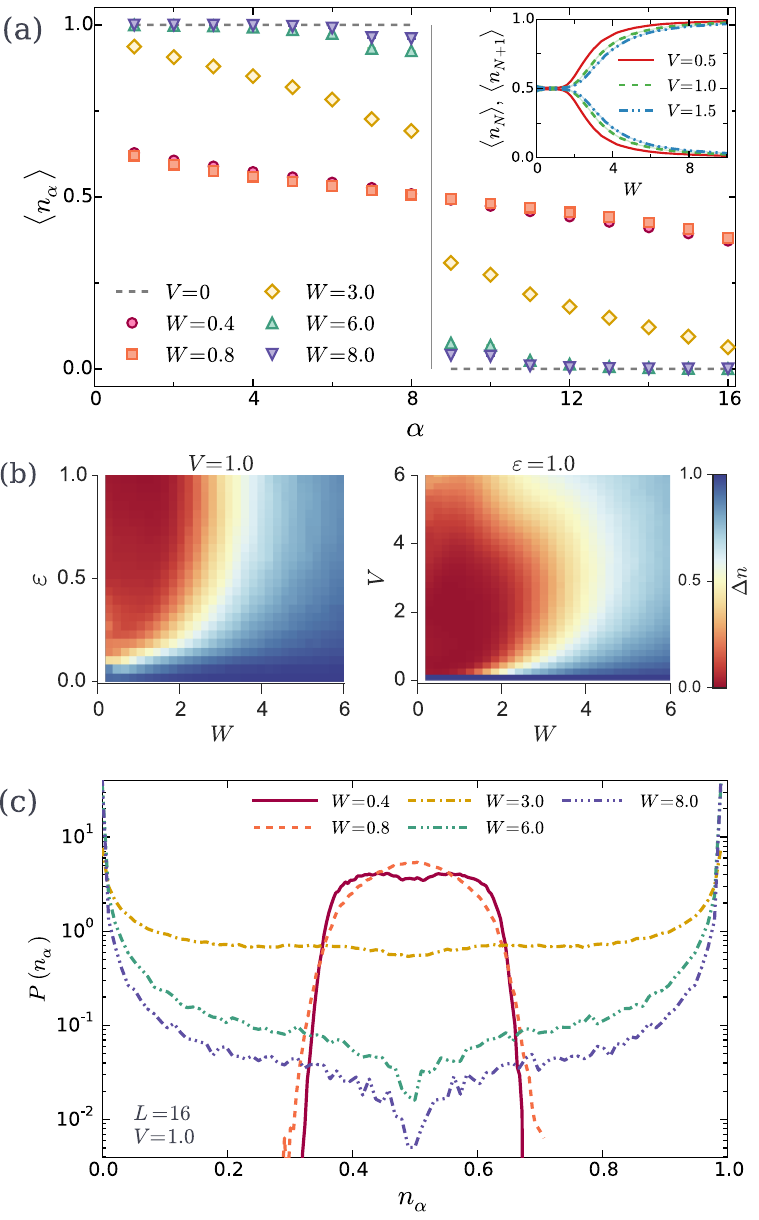}
\caption{(Color online)
(a) The main panel shows the disorder-averaged occupation of the natural orbitals with index $\alpha$ for different
values of disorder strength (system size $L=16$, interaction strength $V=1$).
The dotted line shows the occupation in a noninteracting system, which is independent of the disorder strength.
The vertical line indicates the expected discontinuous behavior of the occupations in the infinite system-size limit of the MBL phase.
The emergence of this discontinuity is further illustrated in the inset, which shows the average occupations $\la n_{N}\ra$ and $\la n_{N+1}\ra$ as a function of disorder strength, for $L=16$ and three values of the interaction ($V=0.5,1,1.5$).
(b)~Occupation discontinuity $\Delta n$ as a function of disorder strength and energy density (left panel) or interaction strength (right panel).
(c)~Distribution of occupations $n_\alpha$ in the delocalized phase ($W=0.4, 0.8 $), near the MBL
transition ($W=3.0$),  and deep in the localized  phase ($W=6.0, 8.0$), for system size $L=16$ and interaction strength $V=1$.
}\label{Fig1}
\end{figure}
%
{\it Occupation spectrum.}
In a noninteracting fermionic system, barring degeneracies, each many-body eigenstate $|\psi_n\rangle$ can be written as a Slater determinant of $N$ single-particle states.
The occupations from the one-particle density matrix are then fixed to $n_{\alpha}=0$ or $1$, with the natural orbitals with $n_\alpha=1$ spanning the space of the single-particle states used in the Slater determinant.
Occupations departing from $n_{\alpha}=0$ or $ 1$ can therefore be interpreted as a signature of the true many-body nature of the  eigenstates in the interacting system and, thus, as a proxy of the delocalization of such states in Fock space.
We expect this Fock-space delocalization to be pronounced in the metallic phase, while it should be suppressed in the MBL phase \cite{BAA,Mirlin}.

In Fig.~\ref{Fig1}(a) we show the disorder-averaged occupations for different values of disorder for $L=16$, with $\langle . \rangle$ denoting the disorder average.
The horizontal dashed lines represent the occupations $\avgna =1$ for $\alpha\leq N$, $\avgna =0$ for $\alpha>N$ in the noninteracting limit $V=0$, where the system is Anderson localized for any finite disorder strength.
The quasi-discontinuous jump $\Delta n=n_{N}-n_{N+1}=1$ between these values is indicated by a vertical line.
In the interacting system, deep in the localized phase ($W=6,8$), half of the natural orbitals remain almost fully occupied, $\avgna \approx 1$, with the other half being almost unoccupied, $\avgna \approx 0$.
As one decreases the disorder and approaches the transition ($W=3$), more orbitals acquire a finite occupation, while for even smaller disorder, in the delocalized phase ($W=0.4, 0.8$), the occupation of all orbitals approaches the mean filling fraction, $\avgna \approx N/L=1/2$.

The redistribution of occupations with decreasing disorder goes along with a reduction of the step-like behavior quantified by $\Delta n$.
A more detailed view of this aspect is provided by the inset of Fig.~\ref{Fig1}(a), showing the disorder dependence of the occupations $\la n_{N}\ra, \, \la n_{N+1}\ra$ for three values of interaction strength $V={0.5, 1.0, 1.5}$.
In the delocalized phase, both occupations are close to the mean filling fraction, $\avgna \approx N/L=1/2$, while deep in the localized phase they tend to their asymptotic values $\la n_{N}\ra = 1$, $\la n_{N+1}\ra = 0$~\cite{supp}.
The dependence of the discontinuity $\Delta n$ on energy density, shown in the left panel of Fig.~\ref{Fig1}(b), recovers the many-body mobility edge~\cite{BAA,huse13,kjaell14,luitz15}, while at small and large interaction strengths (right panel) the delocalized phase shrinks, consistent with observations from dynamics in the same model~\cite{BarLev:2015co}.
According to these results, the occupation spectrum serves as a reliable indicator of many-body localization.

{\it One-particle occupation entropy.} A well documented aspect of MBL is the appearance of strong fluctuations around the localization-delocalization transition~\cite{kjaell14,Vasseur:2015bh,Agarwal:2015cu,Potter:2015ub,vosk15}.
In terms of the occupations, this is addressed in Fig.~\ref{Fig1}(c), which shows the probability distribution functions $P(\na)$ for different disorder strengths in a semi-log plot.
In the large disorder limit the distribution is bimodal with peaks at $n_{\alpha}=0,1$, with very little weight in the central region between these extremal values.
This bimodal distribution is characteristic of the localized state, in analogy to the noninteracting scenario.
As expected, close to the transition ($W=3$) the distribution is wide, with significant weight across the whole range of occupations.
Finally, in the delocalized phase with low enough disorder the distribution becomes again narrower, but now is concentrated around the filling fraction $N/L=1/2$.

%
\begin{figure}[t]
\includegraphics[width=0.95\columnwidth]{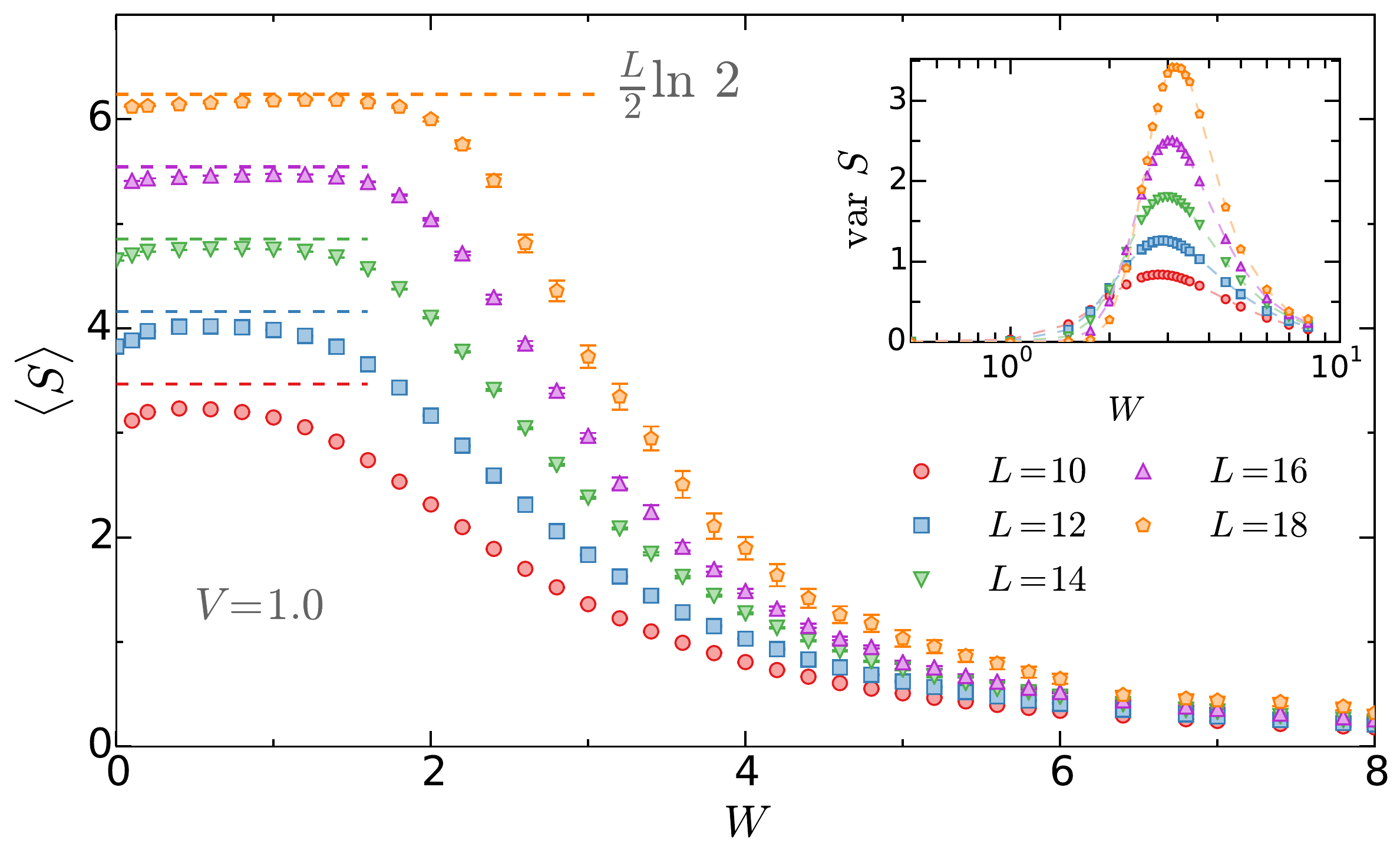}
\caption{(Color online)
Dependence of the disorder averaged one-particle occupation entropy  $\la S\ra$, defined in Eq.~\eqref{eq:entrp}, on the disorder strength, for different system sizes at interaction strength $V=1$.
The dashed lines indicate the maximal value  $\f{L}{2}
\ln2$, corresponding to the volume law for the entropy in a fully delocalized system.
In contrast, in the localized phase the entropy becomes small.
Inset: Variance ${\rm var}\,S$ of the entropy due to sample-to-sample fluctuations in the disorder ensemble as a function of disorder strength, for different system sizes at interaction strength $V=1$.
The peak in the variance indicates the location of the MBL transition.
}\label{Fig2}
\end{figure}
%
In order to quantify these fluctuations further, we consider the entropy
\eq{
  S = -\mathrm{tr} \rho\ln\rho=-\sum_\alpha \na \ln (\na).
  \label{eq:entrp}
}
As this entropy is determined by the occupations of the natural orbitals we call this the one-particle occupation entropy, to distinguish it from the entanglement entropy of the many-body eigenstates.
The disorder-averaged entropy $\la S \ra$ is shown in the main panel of Fig.~\ref{Fig2}, as a function of disorder strength for different system sizes.
In the delocalized phase the entropy approaches the maximal value $\f{L}{2} \ln2$,  indicated by the dashed lines.
This corresponds to a volume law as displayed, in general, by extensive thermodynamic properties and many-body eigenstates in ergodic systems.
In contrast, the entropy in the localized phase is much smaller~\cite{supp}.
%

In the inset of Fig.~\ref{Fig2} we show the variance $\text{var}\, S = \la S^2 \ra - \la S \ra^2$ of the entropy as a function of disorder strength.
For the corresponding case of the entanglement entropy, it is known \cite{Bauer:2013jw,kjaell14} that the variance vanishes in the thermodynamic limit ($L\to\infty$) both in the localized and in the delocalized phase, where in the latter phase this is consistent with the eigenstate thermalization hypothesis.
Furthermore, in finite systems, the variance of the entanglement entropy is sharply peaked in the crossover regime, which is associated with the mixing and coexistence of localized and delocalized regions near the transition, becoming sharper with increasing system size~\cite{kjaell14}.
This universal behavior of the entanglement entropy is mirrored by the one-particle occupation entropy.
The occupation spectrum therefore recovers a reliable signature of the MBL transition, giving quantitative access to the locus of the transition.
%
\begin{figure}[t]
\includegraphics[width=0.95\columnwidth]{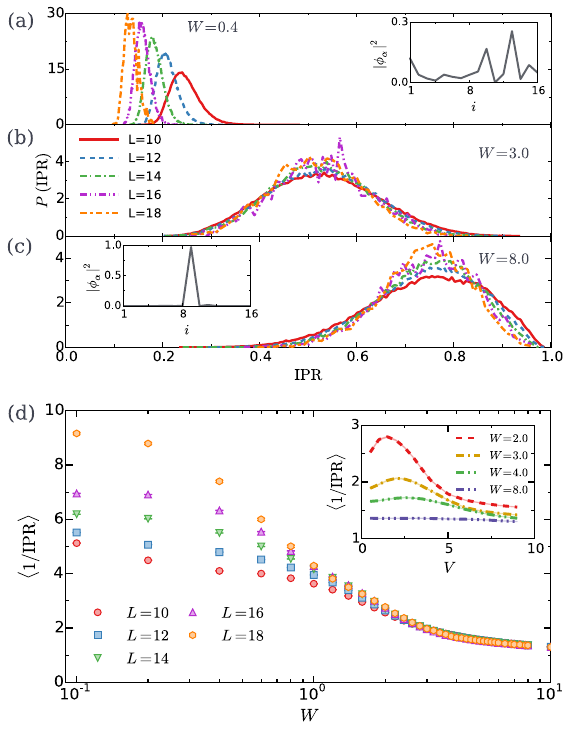}

\caption{(Color online) Evolution of the probability distribution of the IPR for increasing system size
 (a) in the delocalized phase ($W=0.4$), (b) close
to the transition ($W=3.0$), and (c) deep in the localized phase ($W=8.0$).
The insets in (a) and (c) show examples of the natural orbitals.
(d) Average participation ratio $\la 1/\mathrm{IPR}\ra = \xi$ of the natural orbitals as a function of disorder strength.
In the localized phase $\xi$ is independent of the system size, while for small disorder it saturates at $\xi \approx L/2$.
Inset: Average participation ratio as a function of interaction strength $V$ for several values of disorder strength ($L=16$).
}\label{Fig3}
\end{figure}

{\it Delocalization of natural orbitals.} Because of the emerging degeneracy of the occupations deep in the localized and delocalized phase, one may wonder whether the natural orbitals themselves display any signatures of the MBL transition.
As we show in the insets of Figs.~\ref{Fig3}(a) and (c), the orbitals indeed turn out to be well localized in the MBL phase, while they are far more extended in the delocalized phase.
The multiply-peaked structure of the natural orbitals for weak disorder suggests that the delocalization transition involves the formation of a chain throughout the system via which the particle can hop resonantly.
Given the complete set of natural orbitals, a measure of the localization of the occupied states can then be derived from the inverse participation ratio
\eq{
\mathrm{IPR} = \f{1}{N} \sum_{\alpha=1}^L \na \sum_{i=1}^{L} |\phi_\alpha(i)|^4
.
\label{eq:ipr}
}
The IPR is normalized to take the maximal value $1$ for a system in which all occupied states are fully localized, while it takes the minimal value $1/L$ when all occupied states are fully extended.
In between these two extremes, the resonant-hopping picture for the delocalized phase suggests that the IPR scales inversely with the system size, while in the localized phase it should be independent of system size.
These tendencies are confirmed in the main panels of Figs.~\ref{Fig3}(a-c), which show, for three disorder strengths in the delocalized, transitional, and localized regime, how the probability distribution of the IPR depends on the system size.
In the delocalized phase (a), the flow with system size is indicative of a $1/L$ behavior, while in the localized phase (c) the distribution is almost independent of system size, with a peak close to the maximal value $\mathrm{IPR}=1$.
Close to the transition (b), the IPR distribution is wide, with no discernible trend with system size.
It is therefore suggestive to introduce the characteristic length $\xi=\la 1/\mathrm{IPR}\ra$.
Figure \ref{Fig3}(d) shows the disorder-strength-dependence of $\xi$ for different system sizes.
In the localized regime this characteristic length is independent of system size.
With decreasing disorder strength $\xi$ increases, whereas at very small disorder it approaches the value $\xi\approx L/2$.
While $\xi$ is still small at the transition in the accessible system sizes, the orbitals spread out significantly once one enters into the delocalized phase.
Moreover, as shown in the inset of Fig.~\ref{Fig3}(d), $\xi$ depends non-monotonically on $V$: it first increases as $V$ increases, then takes a maximum at a $W$-dependent value and finally decreases again in the large $V$-limit.
A similar behavior was observed in spectral fluctuations in this model~\cite{BarLev:2015co}.
Our quantity $\xi$ thus captures the delocalizing effect of interactions, both in the delocalized and in the MBL phase.

{\it Summary and outlook.}
In conclusion, the one-particle density matrix uncovers essential many-body aspects of interacting disordered fermions.
Our results suggest that in the thermodynamic limit the one-particle occupation spectrum is continuous in the delocalized phase but develops a finite discontinuity in the localized phase.
The corresponding occupation entropy shares features with the many-body entanglement entropy, one of the principal vehicles for the theoretical characterization of the many-body localization transition.
The delocalization is also observed in the structure of the natural orbitals, which is reflected in a system-size dependent inverse participation ratio.
These findings support the conceptual picture that the many-body localization transition involves delocalization both in Fock space and in real space.
Our approach should therefore apply to a broad range of systems that follow this scenario, which can be further enriched when the particle number is not conserved.
An interesting and timely application of our work would consist in analyzing the one-particle density matrix for the system that was experimentally realized in Ref.~\onlinecite{schreiber15}.

{\it Acknowledgment.} We thank  U. Schneider, R. Singh, and F. A. Wolf for very insightful discussions.
\bibliography{references}

\begin{thebibliography}{49}%
\makeatletter
\providecommand \@ifxundefined [1]{%
 \@ifx{#1\undefined}
}%
\providecommand \@ifnum [1]{%
 \ifnum #1\expandafter \@firstoftwo
 \else \expandafter \@secondoftwo
 \fi
}%
\providecommand \@ifx [1]{%
 \ifx #1\expandafter \@firstoftwo
 \else \expandafter \@secondoftwo
 \fi
}%
\providecommand \natexlab [1]{#1}%
\providecommand \enquote  [1]{``#1''}%
\providecommand \bibnamefont  [1]{#1}%
\providecommand \bibfnamefont [1]{#1}%
\providecommand \citenamefont [1]{#1}%
\providecommand \href@noop [0]{\@secondoftwo}%
\providecommand \href [0]{\begingroup \@sanitize@url \@href}%
\providecommand \@href[1]{\@@startlink{#1}\@@href}%
\providecommand \@@href[1]{\endgroup#1\@@endlink}%
\providecommand \@sanitize@url [0]{\catcode `\\12\catcode `\$12\catcode
  `\&12\catcode `\#12\catcode `\^12\catcode `\_12\catcode `\%12\relax}%
\providecommand \@@startlink[1]{}%
\providecommand \@@endlink[0]{}%
\providecommand \url  [0]{\begingroup\@sanitize@url \@url }%
\providecommand \@url [1]{\endgroup\@href {#1}{\urlprefix }}%
\providecommand \urlprefix  [0]{URL }%
\providecommand \Eprint [0]{\href }%
\providecommand \doibase [0]{http://dx.doi.org/}%
\providecommand \selectlanguage [0]{\@gobble}%
\providecommand \bibinfo  [0]{\@secondoftwo}%
\providecommand \bibfield  [0]{\@secondoftwo}%
\providecommand \translation [1]{[#1]}%
\providecommand \BibitemOpen [0]{}%
\providecommand \bibitemStop [0]{}%
\providecommand \bibitemNoStop [0]{.\EOS\space}%
\providecommand \EOS [0]{\spacefactor3000\relax}%
\providecommand \BibitemShut  [1]{\csname bibitem#1\endcsname}%
\let\auto@bib@innerbib\@empty
\bibitem [{\citenamefont {Kramer}\ and\ \citenamefont
  {MacKinnon}(1993)}]{kramer93}%
  \BibitemOpen
  \bibfield  {author} {\bibinfo {author} {\bibfnamefont {B.}~\bibnamefont
  {Kramer}}\ and\ \bibinfo {author} {\bibfnamefont {A.}~\bibnamefont
  {MacKinnon}},\ }\href@noop {} {\bibfield  {journal} {\bibinfo  {journal}
  {Rep. Prog. Phys.}\ }\textbf {\bibinfo {volume} {56}},\ \bibinfo {pages}
  {1469} (\bibinfo {year} {1993})}\BibitemShut {NoStop}%
\bibitem [{\citenamefont {Evers}\ and\ \citenamefont {Mirlin}(2008)}]{evers08}%
  \BibitemOpen
  \bibfield  {author} {\bibinfo {author} {\bibfnamefont {F.}~\bibnamefont
  {Evers}}\ and\ \bibinfo {author} {\bibfnamefont {A.~D.}\ \bibnamefont
  {Mirlin}},\ }\href@noop {} {\bibfield  {journal} {\bibinfo  {journal} {Rev.
  Mod. Phys.}\ }\textbf {\bibinfo {volume} {80}},\ \bibinfo {pages} {1355}
  (\bibinfo {year} {2008})}\BibitemShut {NoStop}%
\bibitem [{\citenamefont {Basko}\ \emph {et~al.}(2006)\citenamefont {Basko},
  \citenamefont {Aleiner},\ and\ \citenamefont {Altshuler}}]{BAA}%
  \BibitemOpen
  \bibfield  {author} {\bibinfo {author} {\bibfnamefont {D.~M.}\ \bibnamefont
  {Basko}}, \bibinfo {author} {\bibfnamefont {I.~L.}\ \bibnamefont {Aleiner}},
  \ and\ \bibinfo {author} {\bibfnamefont {B.~L.}\ \bibnamefont {Altshuler}},\
  }\href@noop {} {\bibfield  {journal} {\bibinfo  {journal} {Ann.\ Phys.}\
  }\textbf {\bibinfo {volume} {321}},\ \bibinfo {pages} {1126} (\bibinfo {year}
  {2006})}\BibitemShut {NoStop}%
\bibitem [{\citenamefont {Gornyi}\ \emph {et~al.}(2005)\citenamefont {Gornyi},
  \citenamefont {Mirlin},\ and\ \citenamefont {Polyakov}}]{Mirlin}%
  \BibitemOpen
  \bibfield  {author} {\bibinfo {author} {\bibfnamefont {I.~V.}\ \bibnamefont
  {Gornyi}}, \bibinfo {author} {\bibfnamefont {A.~D.}\ \bibnamefont {Mirlin}},
  \ and\ \bibinfo {author} {\bibfnamefont {D.~G.}\ \bibnamefont {Polyakov}},\
  }\href@noop {} {\bibfield  {journal} {\bibinfo  {journal} {Phys. Rev. Lett.}\
  }\textbf {\bibinfo {volume} {95}},\ \bibinfo {pages} {206603} (\bibinfo
  {year} {2005})}\BibitemShut {NoStop}%
\bibitem [{\citenamefont {Oganesyan}\ and\ \citenamefont
  {Huse}(2007)}]{oganesyan07}%
  \BibitemOpen
  \bibfield  {author} {\bibinfo {author} {\bibfnamefont {V.}~\bibnamefont
  {Oganesyan}}\ and\ \bibinfo {author} {\bibfnamefont {D.~A.}\ \bibnamefont
  {Huse}},\ }\href@noop {} {\bibfield  {journal} {\bibinfo  {journal} {Phys.
  Rev. B}\ }\textbf {\bibinfo {volume} {75}},\ \bibinfo {pages} {155111}
  (\bibinfo {year} {2007})}\BibitemShut {NoStop}%
\bibitem [{\citenamefont {{\v Z}nidari{\v c}}\ \emph
  {et~al.}(2008)\citenamefont {{\v Z}nidari{\v c}}, \citenamefont {Prosen},\
  and\ \citenamefont {Prelov{\v s}ek}}]{Znidaric:2008cr}%
  \BibitemOpen
  \bibfield  {author} {\bibinfo {author} {\bibfnamefont {M.}~\bibnamefont {{\v
  Z}nidari{\v c}}}, \bibinfo {author} {\bibfnamefont {T.}~\bibnamefont
  {Prosen}}, \ and\ \bibinfo {author} {\bibfnamefont {P.}~\bibnamefont
  {Prelov{\v s}ek}},\ }\href@noop {} {\bibfield  {journal} {\bibinfo  {journal}
  {Phys. Rev. B}\ }\textbf {\bibinfo {volume} {77}},\ \bibinfo {pages} {064426}
  (\bibinfo {year} {2008})}\BibitemShut {NoStop}%
\bibitem [{\citenamefont {Pal}\ and\ \citenamefont {Huse}(2010)}]{pal10}%
  \BibitemOpen
  \bibfield  {author} {\bibinfo {author} {\bibfnamefont {A.}~\bibnamefont
  {Pal}}\ and\ \bibinfo {author} {\bibfnamefont {D.~A.}\ \bibnamefont {Huse}},\
  }\href@noop {} {\bibfield  {journal} {\bibinfo  {journal} {Phys. Rev. B}\
  }\textbf {\bibinfo {volume} {82}},\ \bibinfo {pages} {174411} (\bibinfo
  {year} {2010})}\BibitemShut {NoStop}%
\bibitem [{\citenamefont {Berkelbach}\ and\ \citenamefont
  {Reichman}(2010)}]{Berkelbach:2010ib}%
  \BibitemOpen
  \bibfield  {author} {\bibinfo {author} {\bibfnamefont {T.~C.}\ \bibnamefont
  {Berkelbach}}\ and\ \bibinfo {author} {\bibfnamefont {D.~R.}\ \bibnamefont
  {Reichman}},\ }\href@noop {} {\bibfield  {journal} {\bibinfo  {journal}
  {Phys. Rev. B}\ }\textbf {\bibinfo {volume} {81}},\ \bibinfo {pages} {224429}
  (\bibinfo {year} {2010})}\BibitemShut {NoStop}%
\bibitem [{\citenamefont {Monthus}\ and\ \citenamefont
  {Garel}(2010)}]{Monthus:2010gd}%
  \BibitemOpen
  \bibfield  {author} {\bibinfo {author} {\bibfnamefont {C.}~\bibnamefont
  {Monthus}}\ and\ \bibinfo {author} {\bibfnamefont {T.}~\bibnamefont
  {Garel}},\ }\href@noop {} {\bibfield  {journal} {\bibinfo  {journal} {Phys.
  Rev. B}\ }\textbf {\bibinfo {volume} {81}},\ \bibinfo {pages} {134202}
  (\bibinfo {year} {2010})}\BibitemShut {NoStop}%
\bibitem [{\citenamefont {Bardarson}\ \emph {et~al.}(2012)\citenamefont
  {Bardarson}, \citenamefont {Pollmann},\ and\ \citenamefont
  {Moore}}]{bardarson12}%
  \BibitemOpen
  \bibfield  {author} {\bibinfo {author} {\bibfnamefont {J.~H.}\ \bibnamefont
  {Bardarson}}, \bibinfo {author} {\bibfnamefont {F.}~\bibnamefont {Pollmann}},
  \ and\ \bibinfo {author} {\bibfnamefont {J.~E.}\ \bibnamefont {Moore}},\
  }\href@noop {} {\bibfield  {journal} {\bibinfo  {journal} {Phys. Rev. Lett.}\
  }\textbf {\bibinfo {volume} {109}},\ \bibinfo {pages} {017202} (\bibinfo
  {year} {2012})}\BibitemShut {NoStop}%
\bibitem [{\citenamefont {Huse}\ \emph
  {et~al.}(2013{\natexlab{a}})\citenamefont {Huse}, \citenamefont
  {Nandkishore}, \citenamefont {Oganesyan}, \citenamefont {Pal},\ and\
  \citenamefont {Sondhi}}]{Huse:2013bw}%
  \BibitemOpen
  \bibfield  {author} {\bibinfo {author} {\bibfnamefont {D.~A.}\ \bibnamefont
  {Huse}}, \bibinfo {author} {\bibfnamefont {R.}~\bibnamefont {Nandkishore}},
  \bibinfo {author} {\bibfnamefont {V.}~\bibnamefont {Oganesyan}}, \bibinfo
  {author} {\bibfnamefont {A.}~\bibnamefont {Pal}}, \ and\ \bibinfo {author}
  {\bibfnamefont {S.~L.}\ \bibnamefont {Sondhi}},\ }\href@noop {} {\bibfield
  {journal} {\bibinfo  {journal} {Phys. Rev. B}\ }\textbf {\bibinfo {volume}
  {88}},\ \bibinfo {pages} {014206} (\bibinfo {year}
  {2013}{\natexlab{a}})}\BibitemShut {NoStop}%
\bibitem [{\citenamefont {Serbyn}\ \emph
  {et~al.}(2013{\natexlab{a}})\citenamefont {Serbyn}, \citenamefont
  {Papi{\'c}},\ and\ \citenamefont {Abanin}}]{Serbyn:2013he}%
  \BibitemOpen
  \bibfield  {author} {\bibinfo {author} {\bibfnamefont {M.}~\bibnamefont
  {Serbyn}}, \bibinfo {author} {\bibfnamefont {Z.}~\bibnamefont {Papi{\'c}}}, \
  and\ \bibinfo {author} {\bibfnamefont {D.~A.}\ \bibnamefont {Abanin}},\
  }\href@noop {} {\bibfield  {journal} {\bibinfo  {journal} {Phys. Rev. Lett.}\
  }\textbf {\bibinfo {volume} {110}},\ \bibinfo {pages} {260601} (\bibinfo
  {year} {2013}{\natexlab{a}})}\BibitemShut {NoStop}%
\bibitem [{\citenamefont {Imbrie}()}]{Imbrie:2014vo}%
  \BibitemOpen
  \bibfield  {author} {\bibinfo {author} {\bibfnamefont {J.~Z.}\ \bibnamefont
  {Imbrie}},\ }\href@noop {} {\bibinfo  {journal} {arXiv:1403.7837}\
  }\BibitemShut {NoStop}%
\bibitem [{\citenamefont {Vosk}\ and\ \citenamefont
  {Altman}(2014)}]{Vosk:2014jl}%
  \BibitemOpen
\bibfield  {journal} {  }\bibfield  {author} {\bibinfo {author} {\bibfnamefont
  {R.}~\bibnamefont {Vosk}}\ and\ \bibinfo {author} {\bibfnamefont
  {E.}~\bibnamefont {Altman}},\ }\href@noop {} {\bibfield  {journal} {\bibinfo
  {journal} {Phys. Rev. Lett.}\ }\textbf {\bibinfo {volume} {112}},\ \bibinfo
  {pages} {217204} (\bibinfo {year} {2014})}\BibitemShut {NoStop}%
\bibitem [{\citenamefont {Pekker}\ \emph {et~al.}(2014)\citenamefont {Pekker},
  \citenamefont {Refael}, \citenamefont {Altman}, \citenamefont {Demler},\ and\
  \citenamefont {Oganesyan}}]{Pekker:2014bj}%
  \BibitemOpen
  \bibfield  {author} {\bibinfo {author} {\bibfnamefont {D.}~\bibnamefont
  {Pekker}}, \bibinfo {author} {\bibfnamefont {G.}~\bibnamefont {Refael}},
  \bibinfo {author} {\bibfnamefont {E.}~\bibnamefont {Altman}}, \bibinfo
  {author} {\bibfnamefont {E.}~\bibnamefont {Demler}}, \ and\ \bibinfo {author}
  {\bibfnamefont {V.}~\bibnamefont {Oganesyan}},\ }\href@noop {} {\bibfield
  {journal} {\bibinfo  {journal} {Phys. Rev. X}\ }\textbf {\bibinfo {volume}
  {4}},\ \bibinfo {pages} {011052} (\bibinfo {year} {2014})}\BibitemShut
  {NoStop}%
\bibitem [{\citenamefont {Kj\"all}\ \emph {et~al.}(2014)\citenamefont
  {Kj\"all}, \citenamefont {Bardarson},\ and\ \citenamefont
  {Pollmann}}]{kjaell14}%
  \BibitemOpen
  \bibfield  {author} {\bibinfo {author} {\bibfnamefont {J.~A.}\ \bibnamefont
  {Kj\"all}}, \bibinfo {author} {\bibfnamefont {J.~H.}\ \bibnamefont
  {Bardarson}}, \ and\ \bibinfo {author} {\bibfnamefont {F.}~\bibnamefont
  {Pollmann}},\ }\href@noop {} {\bibfield  {journal} {\bibinfo  {journal}
  {Phys. Rev. Lett.}\ }\textbf {\bibinfo {volume} {113}},\ \bibinfo {pages}
  {107204} (\bibinfo {year} {2014})}\BibitemShut {NoStop}%
\bibitem [{\citenamefont {Laumann}\ \emph {et~al.}(2014)\citenamefont
  {Laumann}, \citenamefont {Pal},\ and\ \citenamefont
  {Scardicchio}}]{Laumann:2014ju}%
  \BibitemOpen
  \bibfield  {author} {\bibinfo {author} {\bibfnamefont {C.~R.}\ \bibnamefont
  {Laumann}}, \bibinfo {author} {\bibfnamefont {A.}~\bibnamefont {Pal}}, \ and\
  \bibinfo {author} {\bibfnamefont {A.}~\bibnamefont {Scardicchio}},\
  }\href@noop {} {\bibfield  {journal} {\bibinfo  {journal} {Phys. Rev. Lett.}\
  }\textbf {\bibinfo {volume} {113}},\ \bibinfo {pages} {200405} (\bibinfo
  {year} {2014})}\BibitemShut {NoStop}%
\bibitem [{\citenamefont {Andraschko}\ \emph {et~al.}(2014)\citenamefont
  {Andraschko}, \citenamefont {Enss},\ and\ \citenamefont
  {Sirker}}]{Andraschko:2014bw}%
  \BibitemOpen
  \bibfield  {author} {\bibinfo {author} {\bibfnamefont {F.}~\bibnamefont
  {Andraschko}}, \bibinfo {author} {\bibfnamefont {T.}~\bibnamefont {Enss}}, \
  and\ \bibinfo {author} {\bibfnamefont {J.}~\bibnamefont {Sirker}},\
  }\href@noop {} {\bibfield  {journal} {\bibinfo  {journal} {Phys. Rev. Lett.}\
  }\textbf {\bibinfo {volume} {113}},\ \bibinfo {pages} {217201} (\bibinfo
  {year} {2014})}\BibitemShut {NoStop}%
\bibitem [{\citenamefont {Chandran}\ \emph {et~al.}(2014)\citenamefont
  {Chandran}, \citenamefont {Khemani}, \citenamefont {Laumann},\ and\
  \citenamefont {Sondhi}}]{chandran14}%
  \BibitemOpen
  \bibfield  {author} {\bibinfo {author} {\bibfnamefont {A.}~\bibnamefont
  {Chandran}}, \bibinfo {author} {\bibfnamefont {V.}~\bibnamefont {Khemani}},
  \bibinfo {author} {\bibfnamefont {C.~R.}\ \bibnamefont {Laumann}}, \ and\
  \bibinfo {author} {\bibfnamefont {S.~L.}\ \bibnamefont {Sondhi}},\
  }\href@noop {} {\bibfield  {journal} {\bibinfo  {journal} {Phys. Rev. B}\
  }\textbf {\bibinfo {volume} {89}},\ \bibinfo {pages} {144201} (\bibinfo
  {year} {2014})}\BibitemShut {NoStop}%
\bibitem [{\citenamefont {De~Luca}\ \emph {et~al.}(2014)\citenamefont
  {De~Luca}, \citenamefont {Altshuler}, \citenamefont {Kravtsov},\ and\
  \citenamefont {Scardicchio}}]{DeLuca:2014ch}%
  \BibitemOpen
  \bibfield  {author} {\bibinfo {author} {\bibfnamefont {A.}~\bibnamefont
  {De~Luca}}, \bibinfo {author} {\bibfnamefont {B.~L.}\ \bibnamefont
  {Altshuler}}, \bibinfo {author} {\bibfnamefont {V.~E.}\ \bibnamefont
  {Kravtsov}}, \ and\ \bibinfo {author} {\bibfnamefont {A.}~\bibnamefont
  {Scardicchio}},\ }\href@noop {} {\bibfield  {journal} {\bibinfo  {journal}
  {Phys. Rev. Lett.}\ }\textbf {\bibinfo {volume} {113}},\ \bibinfo {pages}
  {046806} (\bibinfo {year} {2014})}\BibitemShut {NoStop}%
\bibitem [{\citenamefont {Lazarides}\ \emph {et~al.}()\citenamefont
  {Lazarides}, \citenamefont {Das},\ and\ \citenamefont
  {Moessner}}]{Lazarides2014}%
  \BibitemOpen
  \bibfield  {author} {\bibinfo {author} {\bibfnamefont {A.}~\bibnamefont
  {Lazarides}}, \bibinfo {author} {\bibfnamefont {A.}~\bibnamefont {Das}}, \
  and\ \bibinfo {author} {\bibfnamefont {R.}~\bibnamefont {Moessner}},\
  }\href@noop {} {\bibinfo  {journal} {arXiv:1410.3455}\ }\BibitemShut
  {NoStop}%
\bibitem [{\citenamefont {Ponte}\ \emph {et~al.}(2015)\citenamefont {Ponte},
  \citenamefont {Papi{\'c}}, \citenamefont {Huveneers},\ and\ \citenamefont
  {Abanin}}]{Ponte2015}%
  \BibitemOpen
\bibfield  {journal} {  }\bibfield  {author} {\bibinfo {author} {\bibfnamefont
  {P.}~\bibnamefont {Ponte}}, \bibinfo {author} {\bibfnamefont
  {Z.}~\bibnamefont {Papi{\'c}}}, \bibinfo {author} {\bibfnamefont
  {F.}~\bibnamefont {Huveneers}}, \ and\ \bibinfo {author} {\bibfnamefont
  {D.~A.}\ \bibnamefont {Abanin}},\ }\href@noop {} {\bibfield  {journal}
  {\bibinfo  {journal} {Phys. Rev. Lett.}\ }\textbf {\bibinfo {volume} {114}},\
  \bibinfo {pages} {140401} (\bibinfo {year} {2015})}\BibitemShut {NoStop}%
\bibitem [{\citenamefont {Agarwal}\ \emph {et~al.}(2015)\citenamefont
  {Agarwal}, \citenamefont {Gopalakrishnan}, \citenamefont {Knap},
  \citenamefont {M{\"u}ller},\ and\ \citenamefont {Demler}}]{Agarwal:2015cu}%
  \BibitemOpen
  \bibfield  {author} {\bibinfo {author} {\bibfnamefont {K.}~\bibnamefont
  {Agarwal}}, \bibinfo {author} {\bibfnamefont {S.}~\bibnamefont
  {Gopalakrishnan}}, \bibinfo {author} {\bibfnamefont {M.}~\bibnamefont
  {Knap}}, \bibinfo {author} {\bibfnamefont {M.}~\bibnamefont {M{\"u}ller}}, \
  and\ \bibinfo {author} {\bibfnamefont {E.}~\bibnamefont {Demler}},\
  }\href@noop {} {\bibfield  {journal} {\bibinfo  {journal} {Phys. Rev. Lett.}\
  }\textbf {\bibinfo {volume} {114}},\ \bibinfo {pages} {160401} (\bibinfo
  {year} {2015})}\BibitemShut {NoStop}%
\bibitem [{\citenamefont {Vosk}\ and\ \citenamefont {Altman}(2015)}]{altman14}%
  \BibitemOpen
  \bibfield  {author} {\bibinfo {author} {\bibfnamefont {R.}~\bibnamefont
  {Vosk}}\ and\ \bibinfo {author} {\bibfnamefont {E.}~\bibnamefont {Altman}},\
  }\href@noop {} {\bibfield  {journal} {\bibinfo  {journal} {Annu. Rev.
  Condens. Matter Phys.}\ }\textbf {\bibinfo {volume} {6}},\ \bibinfo {pages}
  {383} (\bibinfo {year} {2015})}\BibitemShut {NoStop}%
\bibitem [{\citenamefont {Nandikishore}\ and\ \citenamefont
  {Huse}(2015)}]{nandkishore15}%
  \BibitemOpen
  \bibfield  {author} {\bibinfo {author} {\bibfnamefont {R.}~\bibnamefont
  {Nandikishore}}\ and\ \bibinfo {author} {\bibfnamefont {D.}~\bibnamefont
  {Huse}},\ }\href@noop {} {\bibfield  {journal} {\bibinfo  {journal} {Annu.
  Rev. Condens. Matter Phys.}\ }\textbf {\bibinfo {volume} {6}},\ \bibinfo
  {pages} {15} (\bibinfo {year} {2015})}\BibitemShut {NoStop}%
\bibitem [{\citenamefont {Serbyn}\ \emph
  {et~al.}(2013{\natexlab{b}})\citenamefont {Serbyn}, \citenamefont
  {Papi\ifmmode~\acute{c}\else \'{c}\fi{}},\ and\ \citenamefont
  {Abanin}}]{serbyn13}%
  \BibitemOpen
  \bibfield  {author} {\bibinfo {author} {\bibfnamefont {M.}~\bibnamefont
  {Serbyn}}, \bibinfo {author} {\bibfnamefont {Z.}~\bibnamefont
  {Papi\ifmmode~\acute{c}\else \'{c}\fi{}}}, \ and\ \bibinfo {author}
  {\bibfnamefont {D.~A.}\ \bibnamefont {Abanin}},\ }\href@noop {} {\bibfield
  {journal} {\bibinfo  {journal} {Phys. Rev. Lett.}\ }\textbf {\bibinfo
  {volume} {111}},\ \bibinfo {pages} {127201} (\bibinfo {year}
  {2013}{\natexlab{b}})}\BibitemShut {NoStop}%
\bibitem [{\citenamefont {Vosk}\ and\ \citenamefont {Altman}(2013)}]{vosk13}%
  \BibitemOpen
  \bibfield  {author} {\bibinfo {author} {\bibfnamefont {R.}~\bibnamefont
  {Vosk}}\ and\ \bibinfo {author} {\bibfnamefont {E.}~\bibnamefont {Altman}},\
  }\href@noop {} {\bibfield  {journal} {\bibinfo  {journal} {Phys. Rev. Lett.}\
  }\textbf {\bibinfo {volume} {110}},\ \bibinfo {pages} {067204} (\bibinfo
  {year} {2013})}\BibitemShut {NoStop}%
\bibitem [{\citenamefont {Huse}\ \emph {et~al.}(2014)\citenamefont {Huse},
  \citenamefont {Nandkishore},\ and\ \citenamefont {Oganesyan}}]{huse14}%
  \BibitemOpen
  \bibfield  {author} {\bibinfo {author} {\bibfnamefont {D.~A.}\ \bibnamefont
  {Huse}}, \bibinfo {author} {\bibfnamefont {R.}~\bibnamefont {Nandkishore}}, \
  and\ \bibinfo {author} {\bibfnamefont {V.}~\bibnamefont {Oganesyan}},\
  }\href@noop {} {\bibfield  {journal} {\bibinfo  {journal} {Phys. Rev. B}\
  }\textbf {\bibinfo {volume} {90}},\ \bibinfo {pages} {174202} (\bibinfo
  {year} {2014})}\BibitemShut {NoStop}%
\bibitem [{\citenamefont {Ros}\ \emph {et~al.}(2015)\citenamefont {Ros},
  \citenamefont {M{\"u}ller},\ and\ \citenamefont {Scardicchio}}]{Ros:2015ib}%
  \BibitemOpen
  \bibfield  {author} {\bibinfo {author} {\bibfnamefont {V.}~\bibnamefont
  {Ros}}, \bibinfo {author} {\bibfnamefont {M.}~\bibnamefont {M{\"u}ller}}, \
  and\ \bibinfo {author} {\bibfnamefont {A.}~\bibnamefont {Scardicchio}},\
  }\href@noop {} {\bibfield  {journal} {\bibinfo  {journal} {Nucl.\ Phys.\ B}\
  }\textbf {\bibinfo {volume} {891}},\ \bibinfo {pages} {420} (\bibinfo {year}
  {2015})}\BibitemShut {NoStop}%
\bibitem [{\citenamefont {Chandran}\ \emph {et~al.}(2015)\citenamefont
  {Chandran}, \citenamefont {Kim}, \citenamefont {Vidal},\ and\ \citenamefont
  {Abanin}}]{Chandran:2015cw}%
  \BibitemOpen
  \bibfield  {author} {\bibinfo {author} {\bibfnamefont {A.}~\bibnamefont
  {Chandran}}, \bibinfo {author} {\bibfnamefont {I.~H.}\ \bibnamefont {Kim}},
  \bibinfo {author} {\bibfnamefont {G.}~\bibnamefont {Vidal}}, \ and\ \bibinfo
  {author} {\bibfnamefont {D.~A.}\ \bibnamefont {Abanin}},\ }\href@noop {}
  {\bibfield  {journal} {\bibinfo  {journal} {Phys. Rev. B}\ }\textbf {\bibinfo
  {volume} {91}},\ \bibinfo {pages} {085425} (\bibinfo {year}
  {2015})}\BibitemShut {NoStop}%
\bibitem [{\citenamefont {Bauer}\ and\ \citenamefont
  {Nayak}(2013)}]{Bauer:2013jw}%
  \BibitemOpen
  \bibfield  {author} {\bibinfo {author} {\bibfnamefont {B.}~\bibnamefont
  {Bauer}}\ and\ \bibinfo {author} {\bibfnamefont {C.}~\bibnamefont {Nayak}},\
  }\href@noop {} {\bibfield  {journal} {\bibinfo  {journal} {J. Stat. Mech.}\
  }\textbf {\bibinfo {volume} {P09005}} (\bibinfo {year} {2013})}\BibitemShut
  {NoStop}%
\bibitem [{\citenamefont {Deutsch}(1991)}]{Deutsch:1991ju}%
  \BibitemOpen
  \bibfield  {author} {\bibinfo {author} {\bibfnamefont {J.~M.}\ \bibnamefont
  {Deutsch}},\ }\href@noop {} {\bibfield  {journal} {\bibinfo  {journal} {Phys.
  Rev. A}\ }\textbf {\bibinfo {volume} {43}},\ \bibinfo {pages} {2046}
  (\bibinfo {year} {1991})}\BibitemShut {NoStop}%
\bibitem [{\citenamefont {Srednicki}(1994)}]{Srednicki:1994dl}%
  \BibitemOpen
  \bibfield  {author} {\bibinfo {author} {\bibfnamefont {M.}~\bibnamefont
  {Srednicki}},\ }\href@noop {} {\bibfield  {journal} {\bibinfo  {journal}
  {Phys. Rev. E}\ }\textbf {\bibinfo {volume} {50}},\ \bibinfo {pages} {888}
  (\bibinfo {year} {1994})}\BibitemShut {NoStop}%
\bibitem [{\citenamefont {Rigol}\ \emph {et~al.}(2008)\citenamefont {Rigol},
  \citenamefont {Dunjko},\ and\ \citenamefont {Olshanii}}]{Rigol:2008bf}%
  \BibitemOpen
  \bibfield  {author} {\bibinfo {author} {\bibfnamefont {M.}~\bibnamefont
  {Rigol}}, \bibinfo {author} {\bibfnamefont {V.}~\bibnamefont {Dunjko}}, \
  and\ \bibinfo {author} {\bibfnamefont {M.}~\bibnamefont {Olshanii}},\
  }\href@noop {} {\bibfield  {journal} {\bibinfo  {journal} {Nature}\ }\textbf
  {\bibinfo {volume} {452}},\ \bibinfo {pages} {854} (\bibinfo {year}
  {2008})}\BibitemShut {NoStop}%
\bibitem [{\citenamefont {Grover}()}]{Grover:2014wm}%
  \BibitemOpen
  \bibfield  {author} {\bibinfo {author} {\bibfnamefont {T.}~\bibnamefont
  {Grover}},\ }\href@noop {} {\bibinfo  {journal} {arXiv:1405.1471}\
  }\BibitemShut {NoStop}%
\bibitem [{\citenamefont {Kondov}\ \emph {et~al.}(2015)\citenamefont {Kondov},
  \citenamefont {McGehee}, \citenamefont {Xu},\ and\ \citenamefont
  {DeMarco}}]{demarco}%
  \BibitemOpen
\bibfield  {journal} {  }\bibfield  {author} {\bibinfo {author} {\bibfnamefont
  {S.~S.}\ \bibnamefont {Kondov}}, \bibinfo {author} {\bibfnamefont {W.~R.}\
  \bibnamefont {McGehee}}, \bibinfo {author} {\bibfnamefont {W.}~\bibnamefont
  {Xu}}, \ and\ \bibinfo {author} {\bibfnamefont {B.}~\bibnamefont {DeMarco}},\
  }\href@noop {} {\bibfield  {journal} {\bibinfo  {journal} {Phys. Rev. Lett.}\
  }\textbf {\bibinfo {volume} {114}},\ \bibinfo {pages} {083002} (\bibinfo
  {year} {2015})}\BibitemShut {NoStop}%
\bibitem [{\citenamefont {Schreiber}\ \emph {et~al.}()\citenamefont
  {Schreiber}, \citenamefont {Hodgman}, \citenamefont {Bordia}, \citenamefont
  {L\"uschen}, \citenamefont {Fischer}, \citenamefont {Vosk}, \citenamefont
  {Altman}, \citenamefont {Schneider},\ and\ \citenamefont
  {Bloch}}]{schreiber15}%
  \BibitemOpen
  \bibfield  {author} {\bibinfo {author} {\bibfnamefont {M.}~\bibnamefont
  {Schreiber}}, \bibinfo {author} {\bibfnamefont {S.~S.}\ \bibnamefont
  {Hodgman}}, \bibinfo {author} {\bibfnamefont {P.}~\bibnamefont {Bordia}},
  \bibinfo {author} {\bibfnamefont {H.~P.}\ \bibnamefont {L\"uschen}}, \bibinfo
  {author} {\bibfnamefont {M.~H.}\ \bibnamefont {Fischer}}, \bibinfo {author}
  {\bibfnamefont {R.}~\bibnamefont {Vosk}}, \bibinfo {author} {\bibfnamefont
  {E.}~\bibnamefont {Altman}}, \bibinfo {author} {\bibfnamefont
  {U.}~\bibnamefont {Schneider}}, \ and\ \bibinfo {author} {\bibfnamefont
  {I.}~\bibnamefont {Bloch}},\ }\href@noop {} {\bibinfo  {journal}
  {arXiv:1501.05661}\ }\BibitemShut {NoStop}%
\bibitem [{\citenamefont {Penrose}\ and\ \citenamefont
  {Onsager}(1956)}]{onsager}%
  \BibitemOpen
\bibfield  {journal} {  }\bibfield  {author} {\bibinfo {author} {\bibfnamefont
  {O.}~\bibnamefont {Penrose}}\ and\ \bibinfo {author} {\bibfnamefont
  {L.}~\bibnamefont {Onsager}},\ }\href@noop {} {\bibfield  {journal} {\bibinfo
   {journal} {Phys. Rev.}\ }\textbf {\bibinfo {volume} {104}},\ \bibinfo
  {pages} {576} (\bibinfo {year} {1956})}\BibitemShut {NoStop}%
\bibitem [{\citenamefont {Koch}\ and\ \citenamefont {Goedecker}(2001)}]{koch}%
  \BibitemOpen
  \bibfield  {author} {\bibinfo {author} {\bibfnamefont {E.}~\bibnamefont
  {Koch}}\ and\ \bibinfo {author} {\bibfnamefont {S.}~\bibnamefont
  {Goedecker}},\ }\href@noop {} {\bibfield  {journal} {\bibinfo  {journal}
  {Solid State Commun.}\ }\textbf {\bibinfo {volume} {119}},\ \bibinfo {pages}
  {105} (\bibinfo {year} {2001})}\BibitemShut {NoStop}%
\bibitem [{\citenamefont {Nessi}\ and\ \citenamefont {Iucci}(2011)}]{nessi11}%
  \BibitemOpen
  \bibfield  {author} {\bibinfo {author} {\bibfnamefont {N.}~\bibnamefont
  {Nessi}}\ and\ \bibinfo {author} {\bibfnamefont {A.}~\bibnamefont {Iucci}},\
  }\href@noop {} {\bibfield  {journal} {\bibinfo  {journal} {Phys. Rev. A}\
  }\textbf {\bibinfo {volume} {84}},\ \bibinfo {pages} {063614} (\bibinfo
  {year} {2011})}\BibitemShut {NoStop}%
\bibitem [{\citenamefont {Gramsch}\ and\ \citenamefont
  {Rigol}(2012)}]{Gramsch:2012bl}%
  \BibitemOpen
  \bibfield  {author} {\bibinfo {author} {\bibfnamefont {C.}~\bibnamefont
  {Gramsch}}\ and\ \bibinfo {author} {\bibfnamefont {M.}~\bibnamefont
  {Rigol}},\ }\href@noop {} {\bibfield  {journal} {\bibinfo  {journal} {Phys.
  Rev. A}\ }\textbf {\bibinfo {volume} {86}},\ \bibinfo {pages} {053615}
  (\bibinfo {year} {2012})}\BibitemShut {NoStop}%
\bibitem [{\citenamefont {De~Luca}\ and\ \citenamefont
  {Scardicchio}(2013)}]{DeLuca:2013ba}%
  \BibitemOpen
  \bibfield  {author} {\bibinfo {author} {\bibfnamefont {A.}~\bibnamefont
  {De~Luca}}\ and\ \bibinfo {author} {\bibfnamefont {A.}~\bibnamefont
  {Scardicchio}},\ }\href@noop {} {\bibfield  {journal} {\bibinfo  {journal}
  {EPL}\ }\textbf {\bibinfo {volume} {101}},\ \bibinfo {pages} {37003}
  (\bibinfo {year} {2013})}\BibitemShut {NoStop}%
\bibitem [{\citenamefont {Bar~Lev}\ \emph {et~al.}(2015)\citenamefont
  {Bar~Lev}, \citenamefont {Cohen},\ and\ \citenamefont
  {Reichman}}]{BarLev:2015co}%
  \BibitemOpen
  \bibfield  {author} {\bibinfo {author} {\bibfnamefont {Y.}~\bibnamefont
  {Bar~Lev}}, \bibinfo {author} {\bibfnamefont {G.}~\bibnamefont {Cohen}}, \
  and\ \bibinfo {author} {\bibfnamefont {D.~R.}\ \bibnamefont {Reichman}},\
  }\href@noop {} {\bibfield  {journal} {\bibinfo  {journal} {Phys. Rev. Lett.}\
  }\textbf {\bibinfo {volume} {114}},\ \bibinfo {pages} {100601} (\bibinfo
  {year} {2015})}\BibitemShut {NoStop}%
\bibitem [{\citenamefont {Luitz}\ \emph {et~al.}(2015)\citenamefont {Luitz},
  \citenamefont {Laflorencie},\ and\ \citenamefont {Alet}}]{luitz15}%
  \BibitemOpen
  \bibfield  {author} {\bibinfo {author} {\bibfnamefont {D.~J.}\ \bibnamefont
  {Luitz}}, \bibinfo {author} {\bibfnamefont {N.}~\bibnamefont {Laflorencie}},
  \ and\ \bibinfo {author} {\bibfnamefont {F.}~\bibnamefont {Alet}},\
  }\href@noop {} {\bibfield  {journal} {\bibinfo  {journal} {Phys. Rev. B}\
  }\textbf {\bibinfo {volume} {91}},\ \bibinfo {pages} {081103} (\bibinfo
  {year} {2015})}\BibitemShut {NoStop}%
\bibitem [{sup()}]{supp}%
  \BibitemOpen
  \href@noop {} {}\bibinfo {note} {See appendices for finite-size dependence of
  the occupation spectrum and the associated occupation entropy.}\BibitemShut
  {Stop}%
\bibitem [{\citenamefont {Huse}\ \emph
  {et~al.}(2013{\natexlab{b}})\citenamefont {Huse}, \citenamefont
  {Nandkishore}, \citenamefont {Oganesyan}, \citenamefont {Pal},\ and\
  \citenamefont {Sondhi}}]{huse13}%
  \BibitemOpen
  \bibfield  {author} {\bibinfo {author} {\bibfnamefont {D.~A.}\ \bibnamefont
  {Huse}}, \bibinfo {author} {\bibfnamefont {R.}~\bibnamefont {Nandkishore}},
  \bibinfo {author} {\bibfnamefont {V.}~\bibnamefont {Oganesyan}}, \bibinfo
  {author} {\bibfnamefont {A.}~\bibnamefont {Pal}}, \ and\ \bibinfo {author}
  {\bibfnamefont {S.~L.}\ \bibnamefont {Sondhi}},\ }\href@noop {} {\bibfield
  {journal} {\bibinfo  {journal} {Phys. Rev. B}\ }\textbf {\bibinfo {volume}
  {88}},\ \bibinfo {pages} {014206} (\bibinfo {year}
  {2013}{\natexlab{b}})}\BibitemShut {NoStop}%
\bibitem [{\citenamefont {Vasseur}\ \emph {et~al.}(2015)\citenamefont
  {Vasseur}, \citenamefont {Potter},\ and\ \citenamefont
  {Parameswaran}}]{Vasseur:2015bh}%
  \BibitemOpen
  \bibfield  {author} {\bibinfo {author} {\bibfnamefont {R.}~\bibnamefont
  {Vasseur}}, \bibinfo {author} {\bibfnamefont {A.~C.}\ \bibnamefont {Potter}},
  \ and\ \bibinfo {author} {\bibfnamefont {S.~A.}\ \bibnamefont
  {Parameswaran}},\ }\href@noop {} {\bibfield  {journal} {\bibinfo  {journal}
  {Phys. Rev. Lett.}\ }\textbf {\bibinfo {volume} {114}},\ \bibinfo {pages}
  {217201} (\bibinfo {year} {2015})}\BibitemShut {NoStop}%
\bibitem [{\citenamefont {Potter}\ \emph {et~al.}()\citenamefont {Potter},
  \citenamefont {Vasseur},\ and\ \citenamefont {Parameswaran}}]{Potter:2015ub}%
  \BibitemOpen
  \bibfield  {author} {\bibinfo {author} {\bibfnamefont {A.~C.}\ \bibnamefont
  {Potter}}, \bibinfo {author} {\bibfnamefont {R.}~\bibnamefont {Vasseur}}, \
  and\ \bibinfo {author} {\bibfnamefont {S.~A.}\ \bibnamefont {Parameswaran}},\
  }\href@noop {} {\bibinfo  {journal} {arXiv:1501.03501}\ }\BibitemShut
  {NoStop}%
\bibitem [{\citenamefont {Vosk}\ \emph {et~al.}()\citenamefont {Vosk},
  \citenamefont {Huse},\ and\ \citenamefont {Altman}}]{vosk15}%
  \BibitemOpen
\bibfield  {journal} {  }\bibfield  {author} {\bibinfo {author} {\bibfnamefont
  {R.}~\bibnamefont {Vosk}}, \bibinfo {author} {\bibfnamefont {D.~A.}\
  \bibnamefont {Huse}}, \ and\ \bibinfo {author} {\bibfnamefont
  {E.}~\bibnamefont {Altman}},\ }\href@noop {} {\bibinfo  {journal}
  {arXiv:1412.3117}\ }\BibitemShut {NoStop}%
\end{thebibliography}%

\appendix

\begin{figure*}[tb]
\includegraphics[width=.98\textwidth]{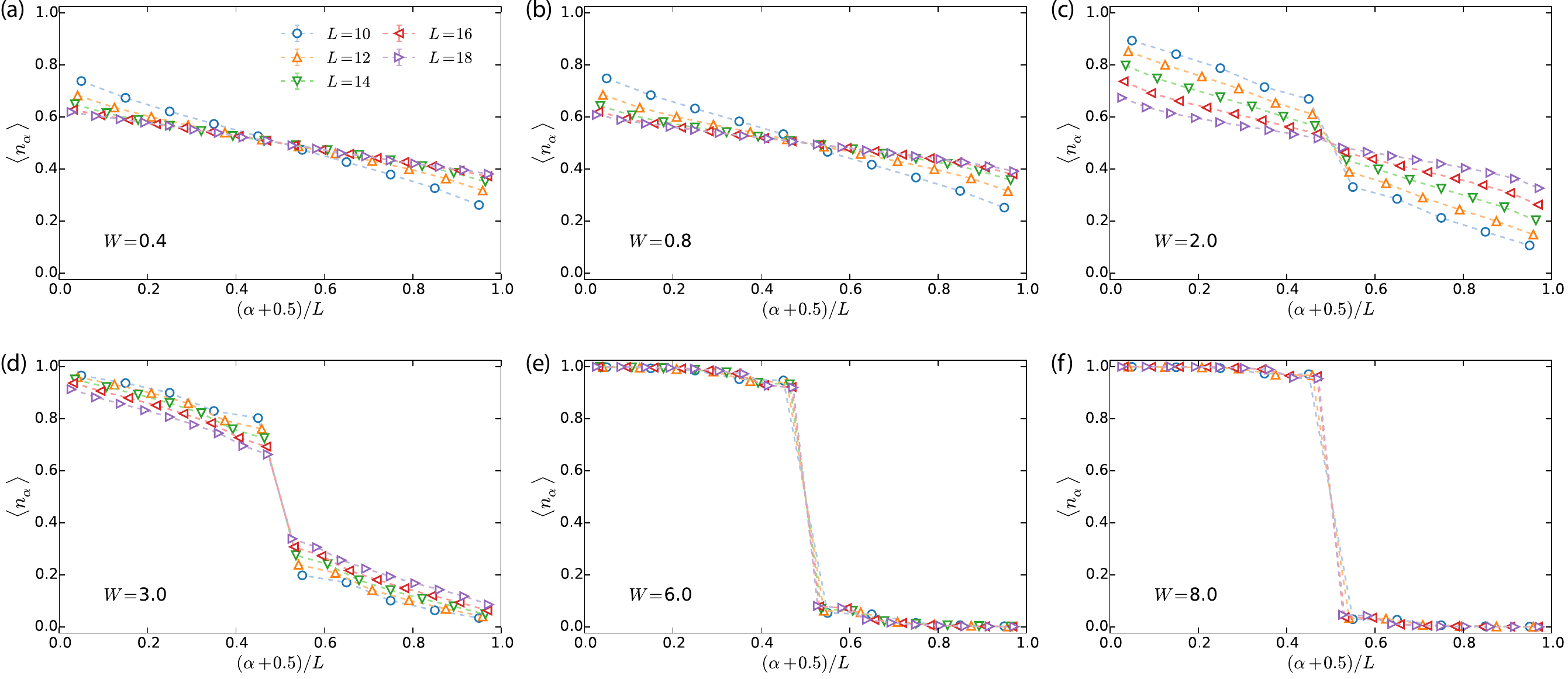}
\caption{\label{figs1}
Occupation spectrum $\langle n_\alpha \rangle$ at fixed disorder strengths  for (a) $W=0.4$, (b) $W=0.8$, (c)  $W=2.0$, (d) $W=3.0$,  (e) $W=6.0$, and  (f) $W=8.0$, for different system sizes $L$ [cf. main panel of Fig.~1(a) in the main text]. The index $\alpha$ is  scaled so that in the thermodynamic limit the horizontal axis runs from 0 to 1.
The interaction strength is $V=1$.
}
\end{figure*}

\begin{figure}[tb]
\noindent
\includegraphics[width=.95\columnwidth]{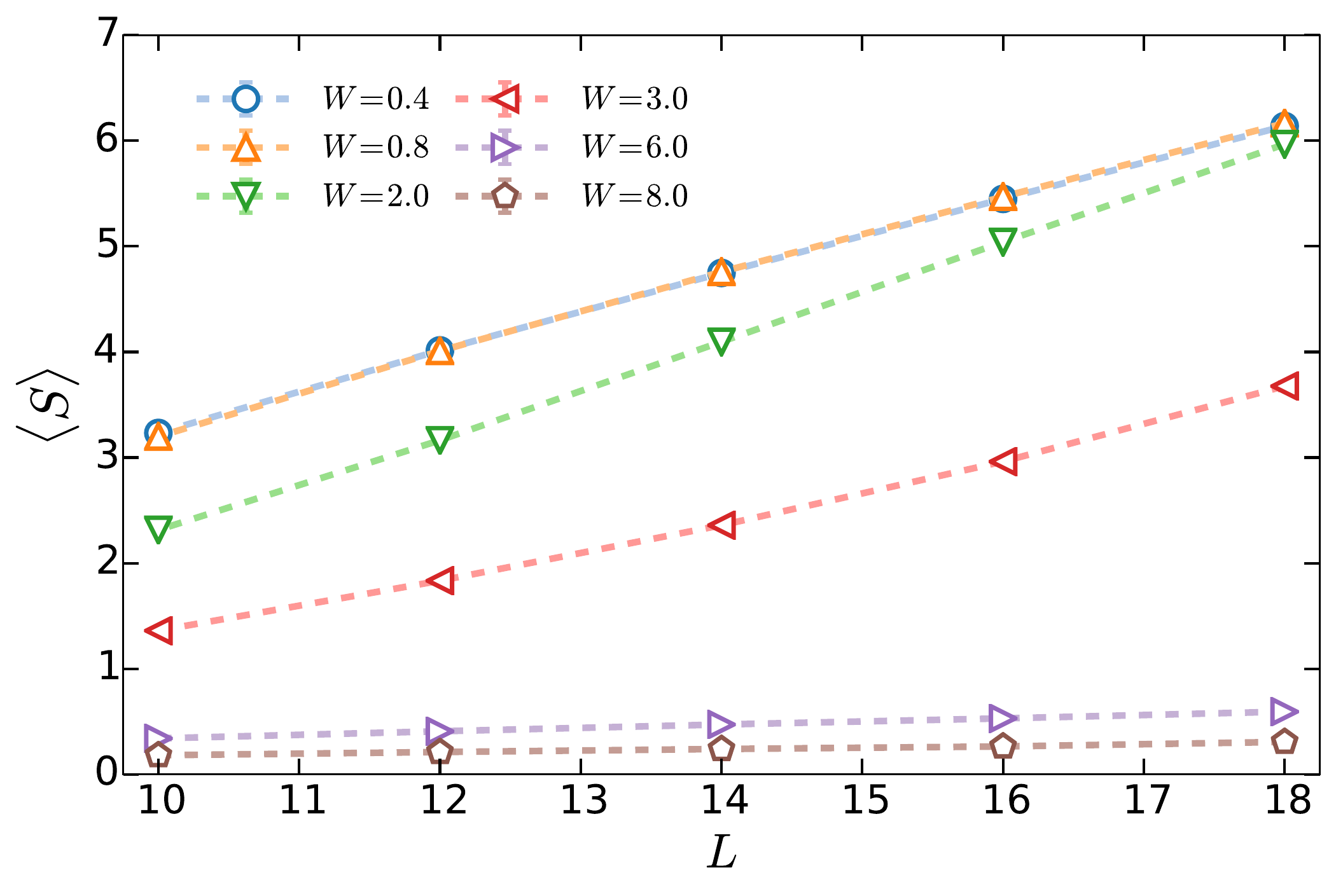}%
\caption{System-size dependence of the occupation entropy $\langle S\rangle$ versus $L$ for disorder strength $W=0.4,0.8,2.0,3.0,6.0,8.0$ with interaction strength $V=1$.\label{figs2}
}
\end{figure}

\section{Appendix A: System-size dependence of the occupation spectrum}

In the main text, we showed that the occupation spectrum $\langle n_\alpha \rangle$ derived from the one-particle density matrix allows to distinguish the many-body localized phase from the delocalized phase. In the localized phase, the occupations tend to be close to 1 and 0, which suggests a discontinuous dependence (finite jump $\Delta n$) to persist in the thermodynamic limit. Likewise, in the delocalized phase, one expects that the eigenstate thermalization hypothesis holds, which corresponds to a smooth behavior of the occupation spectrum in the thermodynamic limit. While the system sizes accessible in exact numerics are still relatively small, Fig.~\ref{figs1} 
shows that these expectations are consistent with our numerical results.
In the delocalized phase [panels (a)-(c)], increasing the system size smoothes out the occupation spectrum (closer to the transition this happens more slowly), while in the localized phase [panels (d)-(f)] we maintain two branches at  occupations close to 1 or 0.
In a more detailed scenario still consistent with this data, the  jump $\Delta n$ would continuously increase from 0 at the MBL transition to 1 deep in the MBL phase. 

\section{Appendix B: System-size dependence of the occupation entropy}

Figure~\ref{figs2} shows the $L$-dependence of the occupation entropy for several values of $L$.
The data is consistent with the overall trend discussed in the main text: $S$ is large
in the delocalized phase and then becomes much smaller in the many-body localized phase, approaching zero for very strong disorder.
In the delocalized phase, the entropy is proportional to $L$ and becomes independent of disorder at large system sizes, consistent with thermalization.
For the accessible system sizes, we observe that the entropy still increases in the localized phase,
which is related to the existence of occupations $n_{\alpha}\not=0,1$.
In fact,  the entanglement entropy also has a residual $L$-dependence for values of $W$ in the localized phase close to the transition \cite{kjaell14}.
As a caution, we stress that the accessible system sizes may be too small to resolve the asymptotic $L$-dependencies of either entropy for values of $W$ sufficiently close to the transition. As long as the localization length is comparable to the system size, we do not expect to be 
able to resolve an area law numerically.

\end{document}